# Integrating Interactive Visualizations in the Search Process of Digital Libraries and IR Systems

Daniel Hienert, Frank Sawitzki, Philipp Schaer, Philipp Mayr

GESIS – Leibniz Institute for the Social Sciences
Lennéstr. 30, 53113 Bonn, Germany
{daniel.hienert, frank.sawitzki, philipp.schaer, philipp.mayr}@gesis.org

**Abstract.** Interactive visualizations for exploring and retrieval have not yet become an integral part of digital libraries and information retrieval systems. We have integrated a set of interactive graphics in a real world social science digital library. These visualizations support the exploration of search queries, results and authors, can filter search results, show trends in the database and can support the creation of new search queries. The use of weighted brushing supports the identification of related metadata for search facets. We discuss some use cases of the combination of IR systems and interactive graphics. In a user study we verify that users can gain insights from statistical graphics intuitively and can adopt interaction techniques.

**Keywords:** Visualization, IR System, Digital Library, Interactivity.

## 1 Introduction

Visualizing information in digital libraries is an ongoing research field in the information retrieval community. But until now visualization has not become an integral part of real world digital libraries (DL) and information retrieval (IR) systems. This may be due to complexity, abstractness and missing interactivity of graphics. Also a lot of systems are stand-alone where the visualization takes most of the space and is not integrated in the standard *search field* - *result list* paradigm. In our approach we have integrated interactive visualizations directly in the search process. This allows the combination of advanced queries, IR features, different information types and interactive graphics. With the ability to filter information items to certain topics, authors etc. and the graphical display of several facet information different research questions can be intuitively answered. Interactivity in visualizations allows to explore information, filter results or start a new search and to explore related metadata. In a user study we show that graphics and interaction techniques can used intuitively and questions can be answered very fast.

In the following sections we give an overview of related work and explain the integration of visualizations in a real world system. Section 4 discusses visualizations and interaction techniques in detail. Section 5 shows use cases in Digital Libraries, followed by a user study. We will conclude in section 7.

## 2 Related Work

An upcoming discipline is the research field of human-computer information retrieval (HCIR), which brings together different aspects of IR and human-computer-interaction to enrich studies and design of IR systems [10]. Actively involving humans into the search process would bring more human intelligence and control into the search process. From the HCIR viewpoint humans should be a central part in the design of IR systems that allow a continuous human control with highly interactive and user-friendly environments. This is associated with retrieval supporting techniques like interactive query expansion, faceted search and navigation, relevance feedback, analytics and visual representation. The visual representation of information is one of the key aspects of HCIR as it shifts complex informations to a more user-accessible form. The features of different visualizations are used to make information quickly processable and to give insights that could not be recognized without the visual processing of data. For example, information of co-authorships can be easier recognized in a network graph than in a textual representation as links between several authors can be seen and processed immediately.

Graphics have been used for different aspects of DL like visualization of metadata, queries, results, documents, co-authors and citations. Visualization types include tables [9][7], time lines [8], graphs [11], networks graphs [6] and topic maps [15]. Most recent tools are BiblioViz [13], PaperLens [9], PaperCube [3] and INVISQUE [16]. Most of these tools have in common that they provide an alternative view and access to a document corpora, but are not integrated in the standard *search field - result list* paradigm of Digital Libraries. Gopubmed [2] is a semantic-based search engine for medical papers. Analog to our solution it offers statistical visualizations in the search results for the actual query. Gopubmed shows facet information like top authors, terms, journal, years, cities and countries in lists with integrated bar charts. Publication counts per year are shown in a bar chart to see the distribution of documents over time. A world map shows location facets and a network graph shows collaboration among top authors. User interactivity is limited to items in the lists; these can be clicked to filter the actual query like a facet.

*Coordinated Views* [14] show the same or related data sets in linked visualizations. By exploiting the advantages of alternative visualization types, data can easier be selected and analyzed. North et al. [12] apply this approach to web-based systems. The technique of *brushing-and-linking* coordinates user actions in one view to other views. For example, selecting a subset of the data in one view highlights it in all views. VisGets [4] is a system for exploring and searching web-based information like RSS-Feed items with the help of interactive graphics. The user interface shows a query form and a result list section, but also interactive graphics like time bar charts, a map and a tag cloud that show facets of the actual search result. Based on the concept of *dynamic queries* [1] users can filter results by adjusting time sliders, zoom and pan on the map and click on keywords in the tag cloud. Result items and graphics are updated instantly to represent the new result set. This way, items can interactively be identified that belong to a specific time span, location and topic. Hovering with the mouse over an item in the graphics or in the result list highlights related attributes in all graphics. The novel technique of *weighted brushing* highlights strongly related attributes with a higher opacity than weakly related ones. This way, items can be identified that are strongly connected. The system is based on database technology and JavaScript/AJAX rather than search engine technology. That is why only simple search is support and IR features like different retrieval models, ranking etc. are missing.

## 3 Embedding Visualizations in a Real World DL

We chose the social science information portal Sowiport[1] as a real-world environment for embedding interactive visualizations. Sowiport integrates literature references, persons, institutions, research projects, events and studies from all areas of the social sciences. It currently contains about 7 million information items from 18 databases, including six databases from ProQuest/CSA. Sowiport is offered in German and English. The technical basis for the search functionality is a SOLR system, an open-source search server based on the Lucene library. Sowiport offers facetted filtering for attributes like information type, database, person, subject and time.

Technical basis for visualizations is a web-based visualization toolkit with the support of various interaction techniques [5]. As a data basis we use facets of the actual search query and different web services. Visualizations have been integrated directly in the search results page[2]. After submitting a search query the result page presents query information, results and facets to filter. Visualizations are positioned directly above the search results in a box that can be folded. The data set for visualizations is directly controlled by the search query and filters. Entered search terms, chosen filters from the search form and chosen facets directly control search results and visualizations. Users can submit very broad queries, can filter down to a certain time span, subject or to a single author. Search results and visualizations are always updated and represent the actual result set.

## 4 Visualizations and Interactivity

### 4.1 Temporal and Spatial Distribution

The temporal and spatial distribution of search results is an interesting aspect to discover i.e. where and when to a specific topic has been published or what databases cover which countries. A line/bar chart for temporal distribution and a map for spatial distribution are visualizations to grasp this information quickly.

The temporal chart shows the distribution of results over the last 50 years. If the search results covers less than 15 years we use a bar chart, for more we use a line chart. By using the facet *information type* the chart can be filtered i.e. to documents, journals or research projects. In addition the facet *location* can be used to see the distribution in different countries. The graph is interactive; hovering with the mouse over the graph shows the frequency for the actual year (see Fig. 1).

[1] www.gesis.org/sowiport
[2] for an online demo see http://www.gesis.org/sowiport/en/search

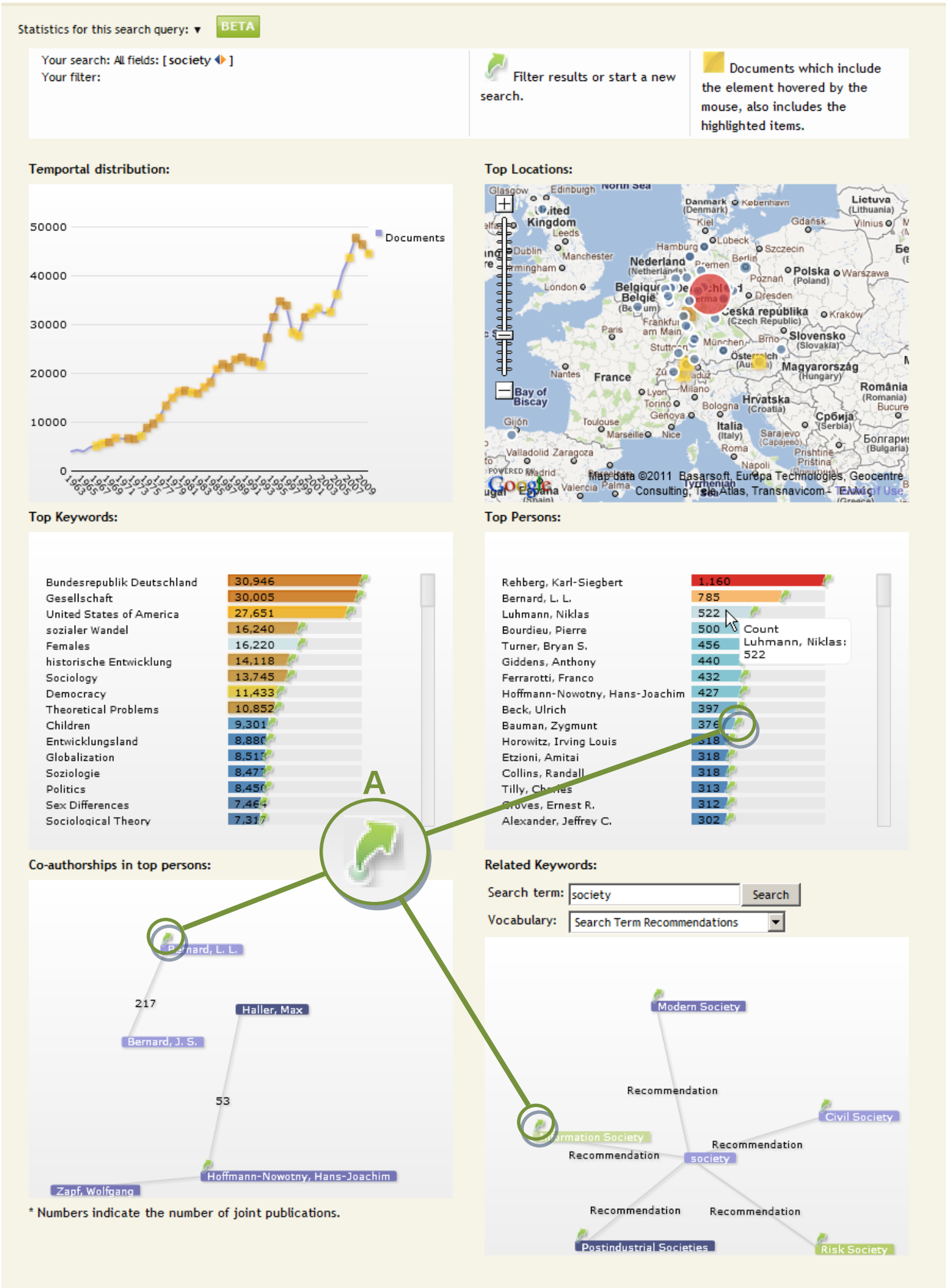

**Fig. 1.** Visualizations in Sowiport: a query for the subject *society* finds 1,066,538 results. The temporal and spatial distribution is shown in the first two graphics, followed by top keywords/top person, co-authorships in top persons and related keywords. The green arrow icon (A) can be used to submit new queries or to filter results. Hovering with the mouse over the person *Luhmann, Niklas* highlights metadata of his information items with times, locations and keywords weighted by occurrence and intensity shown by different yellow to brown color tones.

We also show the distribution of results on a Google map. The map has full interactivity with the possibilities to zoom and pan. Location information is taken from different metadata fields like *publication place* for documents or *place of the institute* for research projects. Location information can differ very much because of different databases, different languages, syntax and abbreviations. We use an internal mapping and a geocoder that resolves this information to geo-coordinates that can be shown on the map. The map's center and zoom level adjusts automatically to the given results.

### 4.2 Information Facets

For every possible facet of a search result we can offer visualizations that show facets frequencies in a horizontal bar chart. For example, the bar chart for the facet *subject* shows up to 50 thesauri terms sorted by frequency. The chart allows the comparison of frequency and importance of certain terms in an easy way.

### 4.3 Co-Authorships and Search Term Recommendation

Two more visualizations base on an analysis of co-authorships for top frequent authors and on several controlled vocabularies including a recommender service to find related search terms. A co-author network for the top 50 authors of a search query is presented in a network graph. The resulting network is presented as a graph with a node for each author and the edge labeled with the count of co-authorships.
Related search terms are presented in a network graph to provide search term recommendations [5]. Data basis can be different vocabulary services the user can choose from: (1) a social science thesaurus, (2) a heterogeneity service, (3) an economy thesaurus and (4) terms from a recommender system based on co-word analysis. Depending on the search query and chosen vocabulary the graph shows related terms and their relations. For example, broader terms, narrower terms, related terms, translations and synonyms are presented for the actual search term.

### 4.4 Searching & Filtering From Visual Items

Any visual item in the graphics can be equipped with a search icon that can initiate a new search or filter the actual results. For example, facets in the top lists like persons or keywords can filter results down to a specific topic or author. Clicking the icon in the co-author graph performs a new search on an author. In the related keywords graph clicking the icon leads to new queries for related or suggested keywords. Figure 1 shows the icon in the different visualizations.

### 4.5 Weighted Brushing

*Brushing-and-linking* is used as a basic technique to highlight related information in different visualizations. Hovering with the mouse over an item in a graphic highlights all connected items in all other graphics. For the linking we use metadata of information items that also contain the element hovered over with the mouse. For example, hovering over an author highlights all associated keywords, times and locations for his information items in the search result (see Fig. 1).

We also use the approach of *weighted brushing* as introduced by VisGets [4]. In their system related visual elements are highlighted according to the count of shared associations. If a visual element A is marked in a graphic, information items (i.e. RSS items) are identified that contain that attribute. If another visual element B shares three information items with A, it is highlighted with a higher opacity than element C which shares only one information item with A. We use the same approach of highlighting related visual items accordingly to their relatedness to the actual selected item. But in contrast to VisGets we did not use a database technology but a highly scalable search engine to retrieve information items and to calculate the linking.

Result sets in an information retrieval system can be very big. If a user searches for a broad topic, results set can contain up to several million items. To identify all possible connections for visualization can be very time consuming in a real time system. Therefore the first step must be to filter the actual result set to a subset whose elements are shown in the graphics. In our case we filter the actual result set to documents that contain the top ten persons and top ten keywords. The SOLR query to retrieve the subset of information items for creating the linking is the following:

```
solr_query = actual_user_query AND person_query AND keyword_query
```

whereby

```
person_query = (person_1 OR person_2 OR … OR person_10)
keyword_query = (keyword_1 OR keyword_2 OR … OR keyword_10)
```

From the resulting set we extract for each document the 4-tuple *{person, keyword, location, time}* and build all permutations of tuples like *{person, keyword}*, *{person, location}* etc. In a second step, we determine the frequency for each tuple. Tuples are used to create the linking and the frequency for the intensity. A resulting linking for a visual item person to keyword maybe look like this:

```
{person, keyword|intensity:123}
```

This means that 123 information items from the result list share the same person and keyword. As a final step the intensity is normalized to values between zero and five so that related visual elements are highlighted according to their relative frequency. The intensity of a connection in the graphics is shown with different color shadings from yellow to brown tones. Yellow highlighted items share less information items than brown colored items.

## 5 Use Cases in Digital Libraries

The combination of a Digital Library with an integrated IR system, different information types and interactive graphics can support answering research questions and transport information faster than before. Visualizations encode information in different ways and give insights that could not be retrieved from a traditional result list. The possibility of creating advanced queries over different metadata fields adds further possibilities to get novel information. In this section we discuss different kinds of research questions by the example of the temporal and spatial distribution that could be retrieved with the combination of a traditional IR system, different information types and interactive visualizations.

Analog to other Digital Libraries users in Sowiport can specify complex queries with Boolean operators over different metadata fields like for example keyword, title, person, source, institutions, year and location. They can also choose between different information types like literature, institutions, journals, events, research projects and studies in 18 different databases. Search results can further be filtered with facets like information type, database, person, subject and time from the result list or from the graphics. This allows the user to filter search results to specific topics, information types, persons, locations, times and complex combinations of these attributes.

The *temporal distribution* of information items is shown in a line chart. In combination with different queries it can be used as an indicator for research activity. For example, users can query for broad topics like *sociological theory* and can see if there is any trend upwards or downwards, maxima and minima over the years. Users are further able to filter the query to a specific author, database, country or time period to see if there is a similar trend. This can also be seen for different databases that covers different areas like German or English databases. The analysis is not only possible for documents, but also for research projects or other information types like events or journals.

This way, questions that could be answered with the combination of specifying queries and the use of integrated visualizations are for example:

- When did the topic X appear? Is there a trend upwards? When has there been a maximum output of publications?
- Has the trend been similar in German and English-speaking countries?
- When did research projects for this topic started? What is the time gap between research projects and dissemination started?
- Is the trend similar for a certain top-author or for a specific subcategory?

The same applies for the *spatial distribution* in the map. The metadata field *location* contains different data for different information types. For example, for documents it means the location of the publisher, for events the venue and for research projects the university address. This way, users can find top locations of events, research projects etc. for a specific topic. So, the following questions could be answered with map view:

- Which countries or universities focus on research on topic X? Is there more research on this topic in Europe, the United States or Asia?
- Where are most conferences on a special topic over time or in a special time period?

## 6 User Study

We have conducted a user study to verify the following hypotheses: (1) Can users gain insights from statistical graphics that could not be retrieved from other resources within search and can users utilize these graphics intuitively? (2) Can new interaction techniques be adapted and understood without any instructions to filter results and find related metadata?

## 6.1 Method

We asked students from a university course of information science to carry out certain tasks in our system and after each task to fill out a questionnaire with the found answer, time needed, difficulty level and comments. First, students could make themselves familiar with the search interface in Sowiport for about two minutes. They could submit a search, view results and open the statistics tab to view the statistical visualizations. In the following study users had to submit a query for the keyword *information society* and should then answer 8 questions only with the help of the visualizations and interaction techniques and to write down the answer in the questionnaire. After conducting the tasks the users could assess the usefulness of visualizations in the search process, rate interaction techniques and give comments and criticism.

## 6.2 Demographics

Participated have 19 students from a university course studying *information science* or *information science and engineering*. Six were male, thirteen female with an average age of 25 and a range from 22 to 30. Participants should rate their experience in dealing with scientific information portals with ratings from 1 to 6 (from very good to very bad). The average rating was good (2.47) with a standard derivation of 0.61.

## 6.3 Tasks and Questions

Participants had to submit a query for the keyword *information society* and then had to answer the following questions only with the help of the visualizations:

a. When the most documents for the topic *information society* had been published and how long the upward trend before lasts based on the number of publications?
b. How many documents were published in Spain?
c. What are the three most frequent keywords and persons for this search?
d. Who has published together with Ullrich Bauer and how many papers were published together?
e. What is the broader term of the search term *information society*?
f. What keywords are associated with the person Rudi Schmiede?
g. Which persons use the keyword *internet*?
h. Use interactive elements in the visualizations to filter search results to the person Rainer Kuhlen.

Tasks (a)-(e) could be answered with the help of a single visualization (i.e. only the map), for the tasks (f) and (g) the interaction techniques weighted brushing had to be used and for task (h) the users had to click the arrow icon. Instructions about how to use the new interaction techniques comes only from the legend that explains the arrow and the weighted brushing symbol and from pop-ups of the green arrow icon.
For each question users should write down the answer in a free text box, answer how long the task took, rate how difficult they perceive the task with a five-point scale (2=very easy, 1=easy, 0=normal, -1=difficult, -2=very difficult) and could give comments or suggestions.

At the end users could give an overall impression by answering the following questions:

- Have visualizations been useful to support the search process?
- Could visualizations like the ones tested support you in your personal research?
- Rate the interaction technique to filter search results or start a new search with the green arrow icon. (5-point scale; 2=very helpful, 1=helpful, 0=fair, -1=not helpful,
- -2=not helpful at all)
- Rate the interaction technique to highlight metadata from common documents with a green frame. (5-point scale; similar to above)
- Do you have general suggestions / comments / criticism?

### 6.4 Results

a. When the most documents for the topic information society had been published and how long the upward trend lasts based on the number of publications?

Users could see the temporal distribution of documents for the search query in the line chart and check with a mouse-over the value of the maxima and the length of the upward trend before. All users could answer the question for the maxima. 13 of 19 gave the correct answer for the length of the trend, 6 participants did not give an answer for the trend at all; comments suggested that the question was not specific enough. The average time needed was 43 seconds with a standard derivation of 49 seconds. The difficulty of the task was rated as easy (1).

b. How many documents were published in Spain?

The spatial distribution of documents can be seen in the map. Users had to zoom and pan to Spain and to check with a mouse-over the count of documents in the pop-up. Two participants wrote down wrong counts, maybe they conducted a slightly different search (i.e. searching the keyword *information society* from the keyword field and not from the overall field). 17 participants succeeded this task with an average time of 21 seconds and a rating of easy (1.33).

c. What are the three most frequent keywords and persons for this search?

Frequent keywords and persons are listed in horizontal bar charts. Users just have to read the three top entries and to write them down. All participants succeeded this task with an average time of 15 seconds and a rating of easy (1.53).

d. Who has published together with Ullrich Bauer and how many papers were published together?

Co-authorships are shown in the network graph, with nodes labeled with co-authors and the count of co-authorship visible at the edge. All participants succeeded this task with an average time of 16 seconds and a rating of easy (1.17).

e. What is the broader term of the search term *information society*?

Related terms are shown in a network graph. Users have to look for the label *broader term* at the edge and to read the term from the node. All users succeeded this task with an average time of 23 seconds and a rating of easy (1.05).

f. What keywords are associated with the person Rudi Schmiede?

For this task users have to use the interaction technique *weighted brushing*. Users have to hover with the mouse pointer over the person in the chart for top persons; associated keywords are then highlighted in the top keywords chart. 17 from 19 participants succeeded this task with an average time of 53 seconds and a rating of easy (0.41). Two participants started a new search to get the information.

g. Which persons use the keyword *internet*?

In this task users have to hover with the mouse pointer over the keyword *internet* and related person are highlighted in the top person chart. The two persons who started a new search in the previous task did the same in this task. 17 from 19 succeeded this task with an average time of 19 seconds and a rating of easy (0.84).

h. Use interactive elements in the visualizations to filter search results to the person Rainer Kuhlen.

Users have to find the name in the top persons chart and start a filtered search by clicking on the arrow icon. Two persons failed, 17 from 19 succeeded this task with an average time of 54 seconds and a difficulty rating of normal (0.5).

**Table 1.** Summarized results of the tasks.

| Question | Correct answers (in %) | Average time needed (in seconds) | Difficulty level (2=very easy, 1=easy, 0=normal, -1=difficult, -2=very difficult) |
|---|---|---|---|
| a. | 68 | 43 | 1 |
| b. | 89 | 21 | 1.33 |
| c. | 100 | 15 | 1.53 |
| d. | 100 | 16 | 1.17 |
| e. | 100 | 23 | 1.05 |
| f. | 89 | 53 | 0.41 |
| g. | 89 | 19 | 0.84 |
| h. | 89 | 54 | 0.50 |

After conducting the tasks, users could assess their overall impression of statistical visualizations supporting the search process: 18 from 19 users found visualizations useful to support the search process and also 18 from 19 users found that visualizations could support them in their personal research. Users rated the interaction technique to filter search results or start a new search with the green arrow icon with an average rating of helpful (1.05) and the interaction technique to highlight metadata from common documents with weighted brushing also with helpful (1.00).

Users could also give general suggestions, comments and criticism. Most users gave comments on the design and the color of the graphics: in the graph for temporal distribution the heading and the legend could be improved. In the map to many locations were overlapping, the map was perceived as to small, and adding a search field or listing countries separately could improve finding a specific country. Useless entries like *no entry* should be filtered out from the persons list. Labels in network graphs could be improved by using a different text color and interaction techniques to pan and zoom in the co-author graph should be integrated. Also users complain that no detailed explanations about the interaction techniques had been given and that using conventions would help. There were also suggestions that the highlighting did not remain when leaving the item with the mouse and that green arrow icons are too small.

### 6.5 Summary

Most users could use the graphics intuitively without any instructions to answer questions for maxima, trends, spatial distribution, top keywords and persons and co-authorships for the actual search query. Even novel interaction techniques for digital libraries like weighted brushing and filtering from graphics could be used and understand only from the legend's explanation. Perceived difficulty level was relatively stable for tasks that could be answered with the help of one individual visualization ($\geq$1) and on a lower level for interaction techniques ($\geq$0.4 and $\leq$0.84) triggered by that they were introduced from the scratch. Users found visualizations useful for search in general and for their personal search and rated both interaction techniques as helpful.

## 7 Conclusion

We have presented a set of interactive visualizations embedded directly in a standard search of a DL. The visualizations support the user in filtering results, interpreting temporal and spatial distribution and finding new suggestions for related keywords and authors. Weighted brushing supports the user in getting a quick overview of related authors, keywords, time and places. Users can see the strength of association by different color shadings.

An IR system allows the user to create advanced and complex queries that filter the information corpus to specific topics, persons and different information types like literature, research projects or events. In combination with interactive visualizations it allows the users to get an overview of research activity, central actors, dissemination activity etc. for this subset as an added value to simple search results.

The information reception of graphical information is fast and adaptive and does not need any instructions. Users in our test had very good results in all tasks and in particularly users have solved these tasks very quickly. Each task was solved in less than one minute, although the users never used visualizations in this context before. The perceived difficulty has been rated *easy* for answering questions with individual visualizations, but also *normal* for newly introduced techniques. In general, users could perform tasks well, quick and easily and that is meeting the challenge of HCIR for interactive and user-friendly environments.